\documentstyle[preprint,aps,epsfig]{revtex}

\newcommand{\be}{\begin{eqnarray}}
\newcommand{\ee}{\end{eqnarray}}

 \newcommand {\dperp} {d_{\perp}}
 \newcommand {\pperp} {p_{\perp}}
 \def\beq{\begin{equation}}
  \def\eeq{\end{equation}}
  \def\beqr{\begin{eqnarray}}
  \def\eeqr{\end{eqnarray}}
  
\def\lsim{\mathrel{\rlap{\lower4pt\hbox{\hskip1pt$\sim$}}
    \raise1pt\hbox{$<$}}}                
\def\gsim{\mathrel{\rlap{\lower4pt\hbox{\hskip1pt$\sim$}}
    \raise1pt\hbox{$>$}}}                

\begin{document}
\draft
\title{
Revealing the black-body regime of small-$x$ deep-inelastic scattering through 
final-state signals}
\author{L. Frankfurt$^{1}$, V. Guzey$^{2}$, M. McDermott$^3$, 
and 
M. Strikman$^4$}
\address{$^1$ School of Physics and Astronomy, Tel Aviv University, 69978 
Tel Aviv, Israel}
\address{$^2$Special Research Centre for the Subatomic Structure of Matter (CSSM),\\
 Adelaide University, 5005, Australia.}
\address{$^3$Division of Theoretical Physics, Dept. of Math. Sciences,\\
 University of Liverpool, Liverpool L69 3BX, England.}
\address{$^4$Department of Physics, Pennsylvania State University,University Park, PA 16802, USA}
\date{\today}
\preprint{
\vbox{
\hbox{ADP-01-13/T448}
\hbox{LTH 497}
}}

\maketitle

\begin{abstract} 

We derive the major characteristics of inclusive and diffractive final states in deep-inelastic scattering 
off heavy nuclei for the-high energy (small-$x$) kinematics in which the limit of complete absorption is 
reached for the dominant hadronic fluctuations in the virtual photon (the black-body limit of the process). 
Both the longitudinal and transverse distributions of the leading hadrons are found to be strikingly different 
from the corresponding ones within the leading-twist approximation, and hence provide unambiguous signals for the onset of the black-body limit.

\end{abstract}

The black-body limit (BBL) of deep-inelastic scattering (DIS) from a heavy nucleus 
was first considered by V.~Gribov in 1969 \cite{gribov}, before the discovery of 
perturbative QCD (PQCD). He assumed that at ultrahigh energies
each hadronic configuration in the photon scatters with 
the same strength, and is absorbed by the nucleus. 
If the (virtual)
photon is  expressed as a linear superposition of partonic components 
(characterised by the transverse and longitudinal momenta of the individual 
partons) then the interaction with the target does not mix up these
partonic components, i.e., they are eigenstates of the scattering matrix. 
This implies that the interaction is diagonal for each hadronic 
fluctuation, and features such as transverse sizes, 
invariant masses, and longitudinal momentum distributions are preserved by 
the BBL interaction. Away from the BBL, the diagonality is grossly violated (see, e.g., \cite{gvdm,FGSvdm}). 

The proton structure function, $F^{p}_2 (x,Q^2)$, measured in DIS at DESY $e\,p$ collider (HERA) 
exhibits a rapid rise with the energy of the photon-proton subprocess, $W$,  
at high energies, or small $x = - q^2 / (2 P q)$, 
where $P$ is the proton's momentum, and $q,q^2=-Q^2$ are the photon's momentum and virtuality. 
The standard leading-twist approximation (LTA) of PQCD describes $F^{p}_2$
extremely successfully in terms of a convolution of perturbatively 
calculable hard coefficient functions with
quark and gluon parton distributions, which evolve in scale according to 
the Dokshitzer-Gribov-Lipatov-Altarelli-Parisi
(DGLAP) \cite{dglap} evolution equations. The rise of $F^{p}_2$ is 
translated into a rapid rise of the gluon and sea-quark distributions 
of the proton with $1/x$ at small $x$. To leading-log accuracy in $Q^2$, 
this leads to a numerically large density of partons at small $x$,  
which calls into question the applicability of the LTA at even smaller $x$
(for recent reviews, see \cite{Mueller,RV,ML,fgs}).
This concern 
is best understood in the target rest frame by considering the inelastic interaction cross sections for given partonic
configurations of the virtual photon. In particular, the expression for 
the scattering of a small $q {\bar q}$ dipole of transverse diameter 
$\dperp$ at small $x$ is given, in terms of the leading-log gluon distribution 
of the nuclear target $xg^A$, by: 
\begin{equation}
{\hat \sigma}_{\mbox{{\scriptsize PQCD}}}^{\mbox{{\scriptsize inel}}}(\dperp^2,x) = \frac{\pi^2}{3} \,
\dperp^2 \, \alpha_s
({\bar Q^2}) \, x g^A (x,{\bar Q^2}) \, , 
\label{eqshat}
\end{equation}
in which the scale ${\bar Q^2} = \lambda/\dperp^2$, with $\lambda$ a 
logarithmic function of $Q^2$. For large gluon densities, this 
leads to a conflict with unitarity of the $S$ matrix for the 
interaction of spatially small wave packets in the virtual photon 
which are almost completely absorbed \cite{FKS} (especially in the case of heavy nuclei, 
where the gluon density is enhanced by nucleon number $A$).
This may be conveniently expressed in terms of 
the geometrical limit for the inelastic ``small dipole - nucleus'' 
interaction of 
\begin{eqnarray}
{\hat{\sigma}}_{\mbox{{\scriptsize PQCD}}}^{\mbox{{\scriptsize inel}}}
&\le {\hat \sigma}_{\mbox{\small{black}}} 
&=  \pi R_{\mbox{{\small target}}}^2 \, , 
\label{sigblack}
\end{eqnarray}
corresponding to the BBL for the total cross section of 
$2\pi R_{\mbox{{\small target}}}^2$.

Numerical studies performed using Eqs.~(\ref{eqshat} and \ref{sigblack}) have
demonstrated that interactions at the upper range of HERA energies in
gluon-induced processes may be fairly close to the BBL
in a wide range of impact parameters
(see, e.g.,
\cite{fgs,mfgs}). This is consistent with an analysis \cite{fsdis99} of 
current fits to HERA diffractive data, which lead to diffractive gluon 
densities which are significantly larger than the quark ones 
(see also \cite{H1dijet}).

For heavy nuclei and central impact parameters, the perturbative $x,Q^2$ region, in which the interaction should be close to the BBL in both quark and
gluon channels, is much broader than for nucleons,
where scattering amplitudes at peripheral impact parameters are important and are far from the BBL.
 For example, for DIS on Pb at $x \sim 10^{-3}$, at small impact parameters the BBL may be reached 
for $Q^2\le 7(15) $GeV$^2$ for the quark (gluon) channel
, that is, for the region where $\ln Q^2/\Lambda_{QCD}^2 \ge \ln x_0/x$ 
($x_0 \sim 0.05 \div 0.1$ is the starting scale 
of the evolution 
at small $x$).
This indicates that DGLAP becomes inapplicable in an $x,Q^2$ range, where the expansion of a small dipole in the transverse 
direction (diffusion)  due to gluon bremsstrahlung  is still small and 
hence the coupling constant is small. This indicates an exciting possibility of the existence of the BBL in QCD in
the perturbative domain for the kinematics which will be probed at CERN Large Hadron Collider (LHC),
and which could also be probed at Relativistic-Heavy-Ion-Collider (RHIC) (at BNL) in pA collisions and at HERA/THERA
in eA mode.

The experimental and theoretical challenge is to find unambiguous 
signals for the BBL. One problem is that predictions for the inclusive 
structure functions do not contain ``smoking gun'' signatures since 
they must account for
leading twist shadowing, uncertainties 
in the input gluon densities, etc.
This is especially acute in the case of the nucleon scattering where, 
even in the BBL, the structure function $F^{p}_{2}(x,Q^2)$
is expected to increase rather rapidly,  
$F^{p}_2 \propto \ln^3 1/x$, of which $\ln^2 1/x$ is due to a Froissart-like 
increase of the radius of interaction $r_{\mbox{{\small int}}} \propto \ln 1/x$ ($\alpha^{\prime}_{{\rm eff}}\propto \ln 1/x$)    and 
the remaining  $\ln 1/x$ reflects the infinite renormalization of the 
electric charge \cite{gribov}.
As discussed in \cite{mfgs}, the latter results 
from the fact that the photon couples in a pointlike manner to quarks. 
This also implies that for any energy there will always be small 
configurations in the photon wave function which have not yet reached the BBL.
However, we expect that the BBL formulae will give a reasonable
description of structure functions and hard diffractive processes
for heavy nuclear target, where the blackness of the interaction is enhanced 
by the number of nucleons. The growth in the radius is essentially screened 
by the nuclear environment so that, for large nuclei, $F^A_2 \propto \ln 1/x$.

To deduce predictions for the properties of the final states it is useful to work within the $q {\bar q}$ 
dipole approximation for the photon wave function. In the BBL the elastic and inelastic contributions are 
equal so that the total cross section is given by twice the inelastic contribution (summed over all dipole configurations):
\begin{equation}
\sigma_{\gamma^*A}= 2 \int {dz~d^2 \pperp \over 2(2\pi)^3}~|\psi_{\gamma^*}^{q {\bar q}} (z, \pperp, Q^2) |^2 \pi R_A^2 \ .
\label{siggama}
\end{equation}
\noindent Here $z$ is the fraction of the photon  plus momentum, $q_+ = q_0 + q_3$, carried by the quark, $z = p_+/q_+$, 
and $\pperp$ is the modulus of its transverse momentum in the plane perpendicular to $\vec{q}$ (the antiquark has $(1-z)$ and $-\pperp$). 
In Eq.~(\ref{siggama}) we implicitly use the fact that the interaction is diagonal in $z$ and $\pperp$ and black for each configuration.
It is convenient to define
\begin{equation}
M^2= (\pperp^2 + m_q^2)/(z(1-z))
\label{msq}
\end{equation}
for the mass squared of the $q\bar q$ system and angle $\theta$ between the direction of
the momentum of the quark in 
the centre of mass frame and the photon direction 
(the transverse plane is defined to be perpendicular to this). Neglecting  $m_q^2$ compared to $M^2, Q^2$ gives 
\begin{equation}
\sin \theta=2 \pperp/\sqrt{M^2}, \quad z= (1 + \cos \theta)/2 \, .
\label{zvar}
\end{equation}
\noindent Hence for the nuclear structure functions we get
\begin{equation}
F^{A}_T (x,Q^2)=\int_0^{M^{2}_{\mbox{{\footnotesize max}}}} dM^2 {2\pi R_A^2\over 12 \pi^3}
{Q^2M^2\rho(M^2)\over (M^2+Q^2)^2} \int_{-1}^1 d \cos \theta ~{3\over 8}(1+\cos^2\theta) \, ,
\label{gr2}
\end{equation}
\begin{equation}  
F_L^{A} (x,Q^2)=\int_0^{M^{2}_{\mbox{{\footnotesize max}}}} dM^2
{2\pi R_A^2\over 12 \pi^3} {Q^4\rho(M^2)\over (M^2+Q^2)^2} \int_{-1}^1 d \cos \theta ~{3\over 4}\sin^2\theta \, ,
\label{gr3}
\end{equation}
\begin{equation}
\rho(M^2)=\sigma^{e^+e^- \to \mbox{{\small hadrons}}}/\sigma^{e^+e^-\to \mu^+\mu^-} \, .
\end{equation}
Integration over $\theta$ gives the transverse and longitudinal 
structure functions in the BBL as an integral over produced masses.
If one takes into account only the suppression induced by the square of the nuclear form factor in the rescattering amplitude, then 
 $M^{2}_{\mbox{{\footnotesize max}}} \le W^2/(m_N R_A)$ and Eqs.~(\ref{gr2}) and (\ref{gr3}) coincide with the original result of \cite{gribov}.
However, in QCD, due to color transparency, $M^{2}_{\mbox{{\footnotesize max}}}$ is determined by the unitarity constraint and is substantially smaller than 
 $W^2/(m_N R_A)$ (for a recent discussion, see \cite{fgs}). The contribution of larger $M^2$ is suppressed by a smaller interaction cross section ($\propto 1/M^2$, in the LTA with small enough gluon densities).
In order to visualize manifestations of the BBL, in the following discussion we neglect this contribution, which has a stronger $A$ dependence than the BBL term.
The inclusion of $\rho(M^2)$ corrects the dipole formula, for which Eq.~(\ref{siggama})
holds, for higher order corrections (in $\alpha_s$) contributing at a given $M^2$. A relation between $\rho(M^2)$ and nuclear structure functions was also suggested in \cite{BP69}.

We also obtain, for the ratio of the deeply virtual Compton scattering and elastic amplitudes,
\begin{equation}
{M(\gamma^*_TA\to \gamma A)\over
 M(\gamma^*_TA\to \gamma^*_T A)}
={
\int_0^{M^{2}_{\mbox{{\footnotesize max}}}} {dM^2 
\rho(M^2)\over (M^2+Q^2)}\over 
\int_0^{M^{2}_{\mbox{{\footnotesize max}}}} {dM^2 
\rho(M^2)M^2\over (M^2+Q^2)^2}} \, .
\end{equation}

The ability to neglect nondiagonal transitions in the BBL \cite{gribov}
justifies the removal of the integral over the masses 
in Eqs.~(\ref{gr2}) and (\ref{gr3}) and, since diffraction is $50\%$ of the total cross section, we 
immediately find for the spectrum of diffractive masses: 
\begin{eqnarray}&&
{dF_T^{D(3)}(x,Q^2,M^2)\over d \cos \theta dM^2}
+{\epsilon dF_L^{D(3)}(x,Q^2,M^2)\over d \cos \theta dM^2}=
\label{difftheta}
\\ 
&&
{\pi R_A^2\over 12 \pi^3}
{Q^2\rho(M^2)\over (M^2+Q^2)^2}({3\over 8}M^2(1+\cos^2\theta)
+{3 \over 4} \epsilon~Q^2 \sin^2\theta) \, , \nonumber
\end{eqnarray}
\noindent where $\epsilon$ is the photon polarization.
Thus, the spectrum of hadrons in the centre of mass of the diffractively 
produced system should be the same as 
in $e^+e^-$ annihilation. 
Hence, the dominant diffractively produced final state will have two jets (with fractal substructure) with a distribution in  
the centre of mass emission angle proportional to $1+\cos^2 \theta$ for the transverse case and $\sin^2 \theta $ for the longitudinal case.  
The diffractive cross section, integrated over $\theta$, is obtained from Eq.~(\ref{difftheta}) by removing the ${3\over 8}(1+\cos^2\theta), {3\over 4} \sin^2\theta$
 factors. 
It follows from Eq.~(\ref{difftheta}) that in 
the BBL diffractive production of high $\pperp$ jets 
is $\propto M^2$ (while in the LTA it is $\propto \ln Q^2$) and hence is enhanced:  
$\left<\pperp^2(jet)\right>_T
= 3~M^2/20, \, \, \left<\pperp^2(jet)\right>_L = M^2/5 \,$ .

The relative rate and distribution of jet variables for three 
jet events (originating from $q\bar q g $ configurations) will also be  
the same as in $e^+e^-$ annihilation and hence is given by the standard
expressions for the process $e^+e^- \to q\bar q g$. 
In addition, in the BBL, the production of jets is flavor democratic 
(weighted by quark charges but unrelated to the quark content of the target).

An important advantage of the diffractive BBL signal is that
these features of the diffractive final state should hold for 
$M^2 \leq Q^2_{\mbox{\scriptsize BBL}}$ even for $Q^2 \geq Q^2_{\mbox{\scriptsize BBL}}\gg \Lambda_{\mbox{{\scriptsize QCD}}}^2$
because configurations with transverse momenta 
$\leq Q_{\mbox{\scriptsize BBL}}/2$ are perturbative but still interact in the black regime 
(and correspond to transverse size fluctuations for which the 
interaction is already black).

Another interesting feature of the BBL is the spectrum of leading
hadrons in the photon fragmentation region. 
It is essentially given by Eq.~(\ref{difftheta}).
Since the distributions in $z$ (or equivalently $\theta$) do not depend on $M$,
the jet distribution in $z$ is given by 
\begin{equation}
{d(\sigma_T +\epsilon \sigma_L)\over d z}\propto {M^2\over Q^2}
 {1+(2z-1)^2\over 8} + \epsilon(z-z^2) \, .
\label{zjet}
\end{equation}

Exclusive vector meson production in the BBL corresponds, in a sense, to a resurrection of
the original vector meson dominance model \cite{Sakurai} without off-diagonal
transitions. 
The amplitude for the vector meson-nucleus interaction
is proportion to $2\pi R_A^2$ (since each configuration in
the virtual photon interacts with the same BBL cross section).
This is markedly different from the requirements \cite{FGSvdm} for matching
the generalised vector dominance model (see, e.g., \cite{gvdm})
with QCD in the scaling limit, where the off-diagonal matrix elements
are large and lead to strong cancellations.
We can factorize out the universal black interaction cross section for the dipole interaction 
from the overlap integral between wave functions of the virtual photon and vector mesons to find,
for the dominant electroproduction of vector mesons,
\begin{equation}
{d\sigma^{\gamma^{\ast}_{T} +A\to V+A} \over dt} 
 =
{M_V^2\over Q^2} {d\sigma^{\gamma^{\ast}_{L} +A\to V+A} \over dt}= 
{(2\pi R_A^2)^2\over 16\pi}{3 \Gamma_V M_V^3 \over \alpha
(M_V^2 + Q^2)^2 } \frac{4~\left|J_1(\sqrt{-t}R_A)\right|^2}{-tR_A^2}\, ,
\label{vm}
\end{equation}
where $\Gamma_V$ is the electronic decay width $V\to e^+e^-$,
$\alpha$ is the fine-structure constant.
Thus the parameter-free prediction is that, in the BBL (complete absorption) at large $Q^2$, vector meson production
cross sections have a $1/Q^2$ behaviour. This is in stark contrast to an asymptotic behaviour of
$1/Q^6$ predicted in PQCD \cite{brod} since a factor $1/Q^4$, due to the square of the cross section of interaction of a small dipole with the target (color transparency), disappears in the BBL.

In the LTA, the factorization theorem is valid and leads to a universal (i.e., target-independent) 
spectrum of leading particles for scattering off
partons of the same flavor. Fundamentally, this can be explained by the fact that, in the Breit frame,
the fast parton which is hit by the photon carries practically all of its light-cone momentum ($z \to 1$).
As a result of QCD evolution, this parton acquires virtuality, $\sim Q^2$, and a rather large transverse momentum, $k_t$ (which is still
$\ll Q^2$). So, in PQCD, quarks and gluons emitted in the process of QCD evolution and in the fragmentation of heavily virtual partons
together still carry all the photon momentum. In contrast, in the BBL, 
the leading particles originate from coherent diffraction (peripheral collisions) and central highly inelastic collisions. These contributions come from
 the fragmentation of a highly virtual $q \bar{q}$ pair with similar light-cone fractions of longitudinal momenta and  
large relative transverse momenta [see, e.g., Eqs.~(\ref{difftheta}) and (\ref{zjet})].

The inclusive spectrum of leading hadrons 
can be assumed, neglecting energy losses, 
 as being due to the  
independent fragmentation of quark and antiquark of virtualities $\geq Q^2$, with $z$ and  $\pperp$ distributions
given by Eqs.~(\ref{difftheta}) and  (\ref{zjet}) (cf. diffractive production of jets discussed above).
Note that energy losses of partons calculated in the limit of small nuclear parton densities do not lead to a change of the the $z$ fraction carried by a parton, and hence do not violate the LTA (see, e.g., \cite{BH93}). In the BBL, energy losses may be larger, further suppressing the spectrum as compared to Eq.~(\ref{eq:nuc:black_nuc:dfs:41}).

The independence of fragmentation is justified because large transverse momenta of quarks 
dominate in the photon wave function  [cf. Eqs.~(\ref{msq})--(\ref{gr3})]
and because of the weakness of the final-state interaction between $q$ and $\bar q$, since
$\alpha_s$ is small and the rapidity interval is of the order of 1.
Obviously, this leads to a gross depletion of the leading hadron spectrum as compared to the LTA situation in which leading
hadrons are produced in the fragmentation region of the parton which carries essentially all  momentum of the virtual
photon.
If we neglect gluon emissions in the photon wave function, we find,
for instance, for the 
differential 
multiplicity of leading hadrons, $dN^{\gamma_{T}^{\ast}/h}/dz$, 
produced by transverse virtual photons, in the BBL,
\begin{equation}
{d N^{\gamma_{T}^{\ast}/h}\over
dz}=2
\int_z^1  D^{q/h}(z/y,Q^2){3\over 4}(1+(2y-1)^2)dy \ .
\label{eq:nuc:black_nuc:dfs:41}
\end{equation}
Here $D^{q/h}(z/y,Q^2)$
is the fragmentation function of a quark, with any flavor $q$, into  hadrons.
To simplify Eq.~(\ref{eq:nuc:black_nuc:dfs:41}) we used $D^{u/h}=D^{d/h}$ and neglected 
a rather small  difference between $D^{u/h}$ and $D^{s,c/h}$.

Calculations using Eq.~(\ref{eq:nuc:black_nuc:dfs:41}) show a strong suppression of the
leading hadron spectrum as compared to the LTA predictions (see Fig.~1). 
Moreover we expect a further softening as $Q^2$ increases, resulting from increased 
parton emission in the virtual photon wave function: progressively more configurations contain 
extra hard gluons, each fragmenting independently in the BBL, further amplifying deviations 
from the standard LTA predictions.

Another important signature of the BBL is the hardening
of $\pperp$ distributions with decreasing $x$ (at fixed $Q^2$). 
Hence, an efficient experimental strategy would be to select leading jets in 
the current fragmentation region and examine their $z$ and $\pperp$-dependence as a function of $x$.
Qualitatively, the effect of broadening of  $\pperp$ distributions is similar to the
increase of the $\pperp$ distribution in \cite{Mclerran}, although final states in DIS were not 
discussed in this model.
However, due to the distribution of the available $\pperp$ between the produced hadrons, this effect would be more difficult to observe than the $z$ depletion one, and one would require detailed modelling.

An important advantage of nondiffractive  scattering is the 
ability to select scattering at central impact parameters
(for example, via studies of inelastic interactions as a function of the number
of nucleons produced in the nucleus fragmentation region). 
Such a selection allows the effective thickness of the nucleus to be increased,  
as compared to the inclusive situation, by a factor of $\sim$ 1.5, and hence
allows the BBL to be reached at significantly larger $x$.
An important signature will be a strong depletion of the leading hadron spectrum with centrality (in the BBL) 
and a lack of correlations with centrality (in the LTA).

To summarize, we have demonstrated that the study of final states in DIS off
nuclei would provide several stringent signals for the onset of the BBL. 
In contrast to inclusive observables they will not depend on details of the input 
parton distributions, and hence will be far more reliable.
In the nucleon target case, despite the fact that many photon configurations are far away from the BBL, 
an examination of these final-states observables appears to be a promising way to search for precursors of the BBL.
In particular, looking for a strong depletion of hadrons 
in the current fragmentation region, in combination with the 
detection of particles in the proton fragmentation region, looks promising.

We thank A.H.~Mueller for a useful discussion and 
GIF, ARC, PPARC, and DOE for support. LF and MS thank INT (University of Washington)  for hospitality
during the time this work was completed.

\begin{figure}[h]
\begin{center}
\epsfig{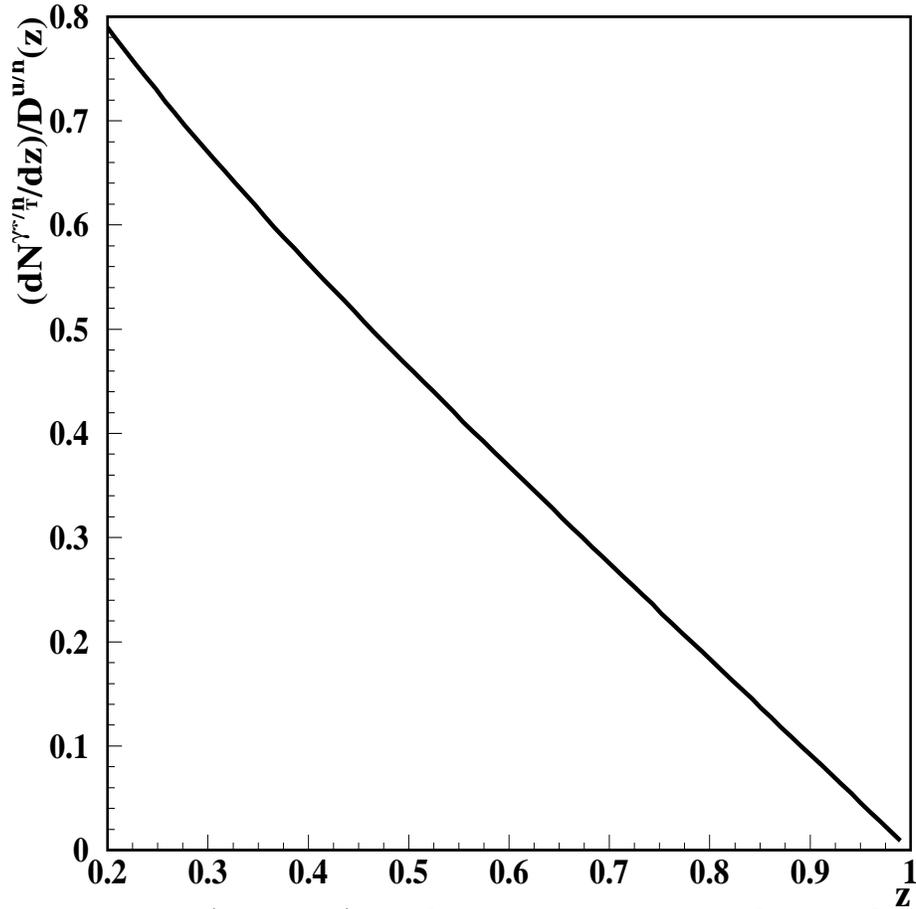}
\caption{The ratio $(d N^{\gamma_{T}^{\ast}/h} /dz)/D^{u/h}(z,Q^2)$
as a function of $z$ at $Q^2$=2 GeV$^2$. $D^{u/h}(z,Q^2)$ is from
\protect\cite{Bourhis}.}
\label{fig:1}
\end{center}
\end{figure}

\end{document}